\documentclass[aps,prl,reprint,superscriptaddress]{revtex4-1}
\usepackage{graphicx}
\usepackage{epstopdf}
\usepackage{dcolumn}
\usepackage{gensymb}
\usepackage{booktabs}
\usepackage{amsmath}
\usepackage{amssymb}
\usepackage{color}
\usepackage{CJK}

\usepackage{bm}        
\usepackage{amssymb,amsmath}   
\usepackage{color}
\emergencystretch=\maxdimen
\begin{document}


\title{Determination of the cluster-decay branching ratio \mbox{from a near-threshold molecular state in $^{10}$Be}}
%


\author{\mbox{W. Jiang}}
\affiliation{School of Physics and State Key Laboratory of Nuclear Physics and Technology, Peking University, Beijing 100871, China}
\affiliation{Institute of High Energy Physics, CAS, Beijing 100049, China}
\affiliation{Spallation Neutron Source Science Center, Dongguan 523803, China}
\author{\mbox{Y.~L. Ye}}
\email{yeyl@pku.edu.cn}
\affiliation{School of Physics and State Key Laboratory of Nuclear Physics and Technology, Peking University, Beijing 100871, China}
\author{\mbox{C.~J. Lin}}
\affiliation{China Institute of Atomic Energy, Beijing 102413, China}
\author{\mbox{Z.~H. Li}}
\author{\mbox{J.~L. Lou}}
\author{\mbox{X.~F. Yang}}
\affiliation{School of Physics and State Key Laboratory of Nuclear Physics and Technology, Peking University, Beijing 100871, China}
\author{\mbox{Q.~T. Li}}
\author{\mbox{Y.~C. Ge}}
\author{\mbox{H. Hua}}
\author{\mbox{D.~X. Jiang}}
\affiliation{School of Physics and State Key Laboratory of Nuclear Physics and Technology, Peking University, Beijing 100871, China}
\author{\mbox{D.~Y. Pang}}
\affiliation{School of Physics and Nuclear Energy Engineering, Beihang University, Beijing 100191, China}
\author{\mbox{J. Li}}
\author{\mbox{J. Chen}}
\affiliation{Physics Division, Argonne National Laboratory, Argonne, Illinois 60439, USA}
\author{\mbox{Z.~H. Yang}}
\affiliation{RCNP, Osaka University, 10-1 Mihogaoka, Ibaraki, Osaka, 567-0047, Japan}
\author{\mbox{X.~H. Sun}}
\affiliation{RIKEN Nishina Center, 2-1 Hirosawa, Wako, Saitama 351-0198, Japan}
\author{\mbox{Z.~Y. Tian}}
\author{\mbox{J. Feng}}
\author{\mbox{B. Yang}}
\author{\mbox{H.~L. Zang}}
\author{\mbox{Q. Liu}}
\author{\mbox{P.~J. Li}}
\author{\mbox{Z.~Q. Chen}}
\author{\mbox{Y. Liu}}
\author{\mbox{Y. Zhang}}
\author{\mbox{J. Ma}}
\affiliation{School of Physics and State Key Laboratory of Nuclear Physics and Technology, Peking University, Beijing 100871, China}
\author{\mbox{H.~M. Jia}}
\author{\mbox{X.~X. Xu}}
\author{\mbox{L. Yang}}
\author{\mbox{N.~R. Ma}}
\author{\mbox{L.~J. Sun}}
\affiliation{China Institute of Atomic Energy, Beijing 102413, China}


\date{\today}
\begin{abstract}

A puzzle has long existed for the $\alpha$-cluster content in the near-threshold 7.54 MeV state of $^{10}$Be. A new measurement was conducted to measure the cluster-decay partial width of this state, using the reaction $\rm{^9Be}(\rm{^9Be}, \rm{^{10}Be}^{*} \rightarrow \alpha + \rm{^6He})\rm{^8Be}$ at 45 MeV beam energy. Special measures were taken to reduce  the strong near-threshold background. The neutron-decay strength was also obtained based on the three-fold coincident measurement. A cluster-decay branching ratio of $(4.04 \pm 1.26)\times 10^{-4}$ is obtained, resulting in a reasonably large $\alpha$-cluster spectroscopic factor. The present work confirms the formation of the $\sigma$-bond molecular rotational band headed by the 6.18 MeV state in $^{10}$Be.

\end{abstract}

\pacs{21.60.Gx, 23.70.+j, 25.70.Hi, 25.70.Ef}



\maketitle

It has been well established that, in light nuclei, the quantum states formed near the cluster-separation threshold tend to possess a large degree of cluster configuration ~\cite{Ikeda1968,Oer2006,Horiuchi2012,Lijing2017,Freer2018}. In recent years clustering has also been identified in neutron-rich unstable nuclei where the excess valence neutrons may act as the valence bonds similar to those in atomic molecular systems ~\cite{Oer2006}.

The persistence of the clustering effect from stable to unstable nuclei has initially been demonstrated along the Beryllium isotopic chain ~\cite{Oer2006,Yang2014,Buck1977,Oer1996}.  In the case of neutron-rich $^{10}$Be, the four excited states around \mbox{6 MeV}  (about 2.5 MeV below the 2n separation threshold) can be perfectly explained by the combination of the valence neutron orbits surrounding the 2-$\alpha$ cores \cite{Oer1996,Oer1997}. These predominant molecular structures in the excited states of $^{10}$Be are also supported by the Antisymmetrized Molecular Dynamics (AMD) calculations \cite{Enyo1999,Lyu2016Be10}, and by various experimental evidences ~\cite{Oer2006}. Based on these speculations, the observed excited states in $^{10}$Be were grouped into several molecular rotational bands headed by states at around \mbox{6 MeV} and characterized by very large moment of inertia \cite{Oer1997,Oer2006}. These molecular bands form a beautiful classification of the excited states in $^{10}$Be ~\cite{Oer2006}. The applicability of this classification depends, of course, on the intrinsic structure of each member state in these bands (\cite{Fortune2011Be10, Aquilla2016, Kra2017, Lyu2018} and references therein).

Among the predicted molecular rotational bands in $^{10}$Be, the one formed by the states at 6.18 MeV ($0_2^+$), 7.54 MeV ($2_3^+$) and ~10.15 MeV ($4_1^+$) has acquired special attention, owing to its pure $\sigma$-bond feature ~\cite{Oer2006,Enyo1999,Lyu2016Be10}, which leads to the longest chain shape corresponding to the largest moment of inertia. In addition to the spin-energy systematics associated with the moment of inertia, the cluster-decay partial width is of essential importance since it determines quantitatively the cluster content in a resonant state.  The cluster-decay strength of the 10.15 MeV state in $^{10}$Be has been measured in many experiments \cite{Fortune2011Be10,Aquilla2016}. A spin-parity of $4^+$ and a $\rm{^6He + \alpha}$ cluster spectroscopic factor (SF) ranging from 0.66 to 2.23, for a channel radius from 1.8 to 1.4 fm, were firmly established, indicating an almost pure cluster structure in this state \cite{Freer2006Be1015,Suzuki2013}. The band head at 6.18 MeV ($0_2^+$) is below the $\alpha$-separation threshold at 7.41 MeV. It is interpreted as a pure $\sigma$-bond molecular state based on its selective population and $\gamma$-decay properties \cite{Oer1996}, and on the consistent theoretical calculations \cite{Oer2006,Lyu2016Be10}. However, for the 7.54 MeV ($2_3^+$) member state, being just 0.132 MeV above the $\alpha$-separation threshold, the experimental investigations related to the cluster-decay are very limited, due mostly to the extreme difficulties in the near-threshold measurement and analysis. The first experimental result was obtained from the reaction $\rm{^7Li}(\mathrm{^7Li}, \mathrm{^{10}Be}^{*} \rightarrow \alpha + \mathrm{^6He})\alpha$ at 34 MeV beam energy ~\cite{Liendo2002}. A branching ratio (BR) of $3.5(12)\times 10^{-3}$ was reported. Another experiment, using the reaction $\rm{^7Li}(\mathrm{^6He}, \mathrm{^{10}Be}^{*} \rightarrow \alpha + \mathrm{^6He})t$ at 18 MeV beam energy, generated only a lower limit of BR  $\ge 2.0(6) \times 10^{-3}$ ~\cite{Milin2005}. We notice that the significantly suppressed $\alpha$-decay fraction is \mbox{basically} attributed to the extremely small relative energy \mbox{against} a large Coulomb barrier. Hence the cluster content inside the mother nucleus may still be large. Using the above referred BR ~\cite{Liendo2002}, an unreasonably large cluster SF of 51(19) for this 7.54 MeV state was extracted by Fortune $et$ $al.$ ~\cite{Fortune2011Be10}. This puzzle has not been solved so far due essentially to the experimental difficulties.

We present here a new measurement of the $\alpha$-decay BR of the 7.54 MeV state in $^{10}$Be, using the reaction $\rm{^9Be}(\rm{^9Be}, \rm{^{10}Be}^{*} \rightarrow \alpha + \rm{^6He})\rm{^8Be}$ at 45 MeV beam \mbox{energy}. This reaction was chosen after a careful  consideration of the beam availability, and the compromise between the optimal detection of the decay fragments and the necessary to avoid the high-flux elastic scattering particles at very forward angles. It would be worth noting that the $\pi$-type orbit in $^9$Be(g.s.) may be replaced by the $\sigma$-type orbit when expanding the distance between the 2-$\alpha$ cores \cite{Oer1996,Oer2006}, which is just the case for the well deformed $^{10}$Be($2_3^+$) state \cite{Enyo1999}. Hence the population of the $\sigma$-bond molecular state, such as the 7.54 MeV ($2^+$) state in $^{10}$Be, is possible in the present one neutron transfer reaction, as justified by some previous experiments \cite{ANDERSON1974,Kuchera2011,Carbone2014,Harakeh1980}. In the present work, some special measures were taken to handle the near-threshold background and to determine the contributions from the neutron-decay channel. Finally a much smaller cluster-BR is obtained for the 7.54 MeV state , being consistent with the theoretical expectations.

The experiment was carried out at the HI-13 tandem accelerator facility at China Institute of Atomic Energy (CIAE) ~\cite{Lijing2017}. A 45 MeV $\rm{^{9}Be}$ beam with an intensity of about 1 pnA was used to bombard a self-supporting $\rm{^{9}Be}$ target (166-$\rm{\mu}$g/cm$^2$). A schematic drawing and detailed descriptions of the detector setup can be found in Refs.\cite{tianzy2016,Lijing2017}. The reaction products were detected and identified by six silicon-strip telescopes, namely U0, U1, U2, D0, D1 and D2, which were placed symmetrically on both sides of the beam axis ~\cite{Lijing2017,Jiang2017,tianzy2016}. Double sided silicon-strip detectors (DSSDs) were employed, providing excellent two-dimensional position resolutions and the ability to record multi-hit events in one telescope. The forward-angle telescope U0(D0) was centered at $23^\circ$ with respect to the beam direction and at a distance of 140 mm from the target.  The large-angle telescopes U1(D1) was at $60^\circ$ and 116 mm, respectively. U2(D2) telescope was installed at the backward angle for other purpose ~\cite{Lijing2017,tianzy2016}. The active area of each telescope is $50 \times 50 ~\rm{mm}^2$. Energy calibration of the Si detectors was realized by using a three-component $\alpha$ source and the elastic scattering of $\rm{^{9}Be}$ off a $\rm{^{197}Au}$ target. The typical energy resolution of the silicon-strip detector was better than 1.0$\%$ for 5.49-MeV $\alpha$ particles \cite{Qiaorui2014, CalibrationMilin2014}. We note that the application of the DSSDs with small-size pixels ($2 \times 2~\rm{mm}^2$ ) is of essential importance here. Since the targeted 7.54 MeV resonance is only 130 keV above the $\alpha$-separation threshold, the opening angle in the laboratory system is small for the decay products, of which the coincident detection efficiency depends sensitively on the pixel size.

\begin{figure}[!b]
\vspace{-4mm}
\begin{center}
\includegraphics[width=.48\textwidth]{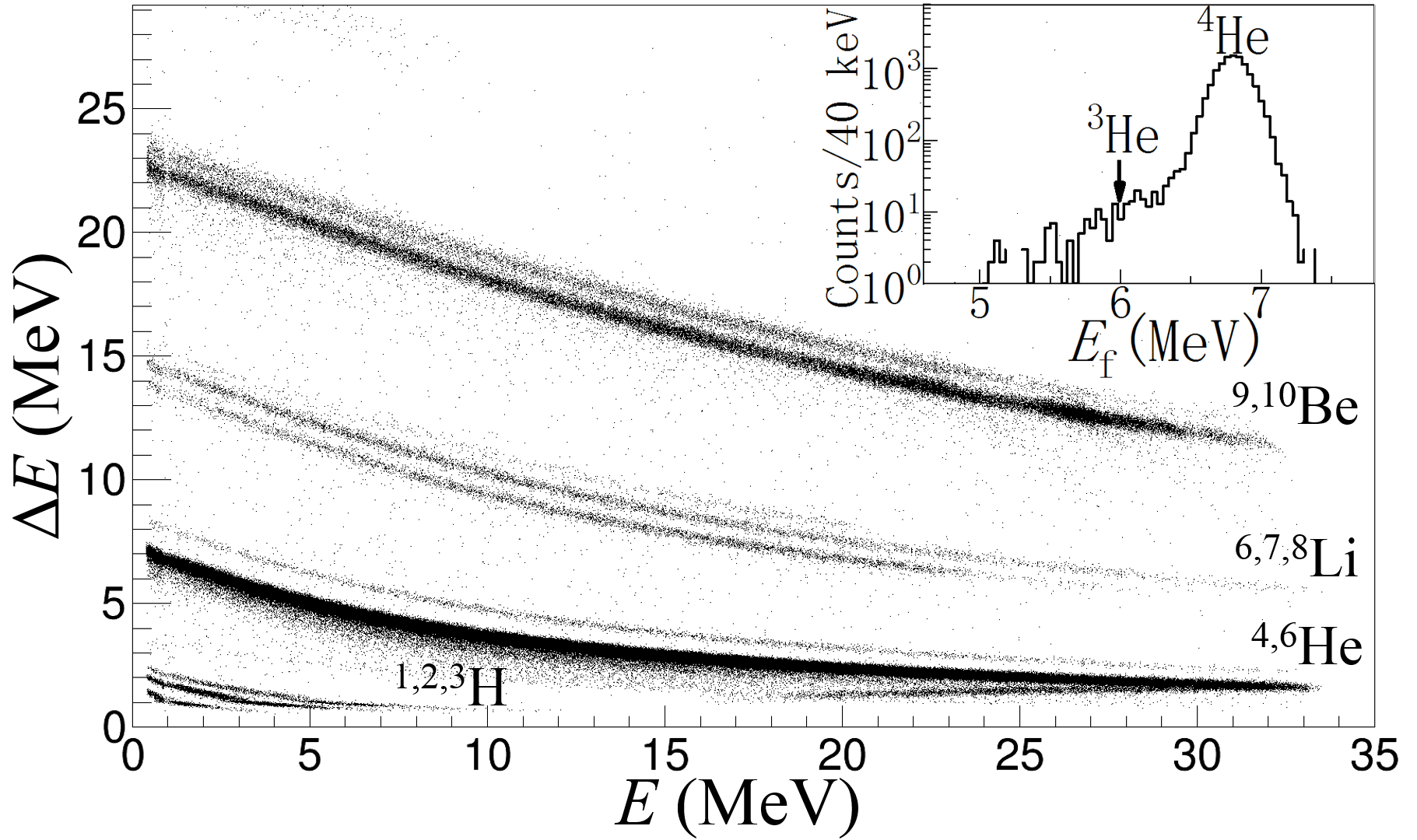}
\vspace{-6mm}
\caption{PID spectrum measured by the U0 telescope using the $\Delta\it{E}$-$E$ method. The inset shows the projected PID spectrum \cite{CalibrationMilin2014} gated on $^6$He as another coincident particle in the same telescope.}
\label{fig:1}
\end{center}
\vspace{-8mm}
\end{figure}

Using the standard energy-loss versus stoping-energy ($\Delta{E}$-$E$) method, excellent particle identification (PID) performance was achieved up to beryllium isotopes (Fig.~\ref{fig:1}). The overall performance of the detection system was checked by reconstructing the $^8$Be energy spectrum from the 2-$\alpha$ particles which were coincidentally detected in one forward-angle telescope \cite{Jiang2017,Freer2004BeB,Aquilla2016}.

\begin{figure}[!b]
\begin{center}
\includegraphics[width=.40\textwidth]{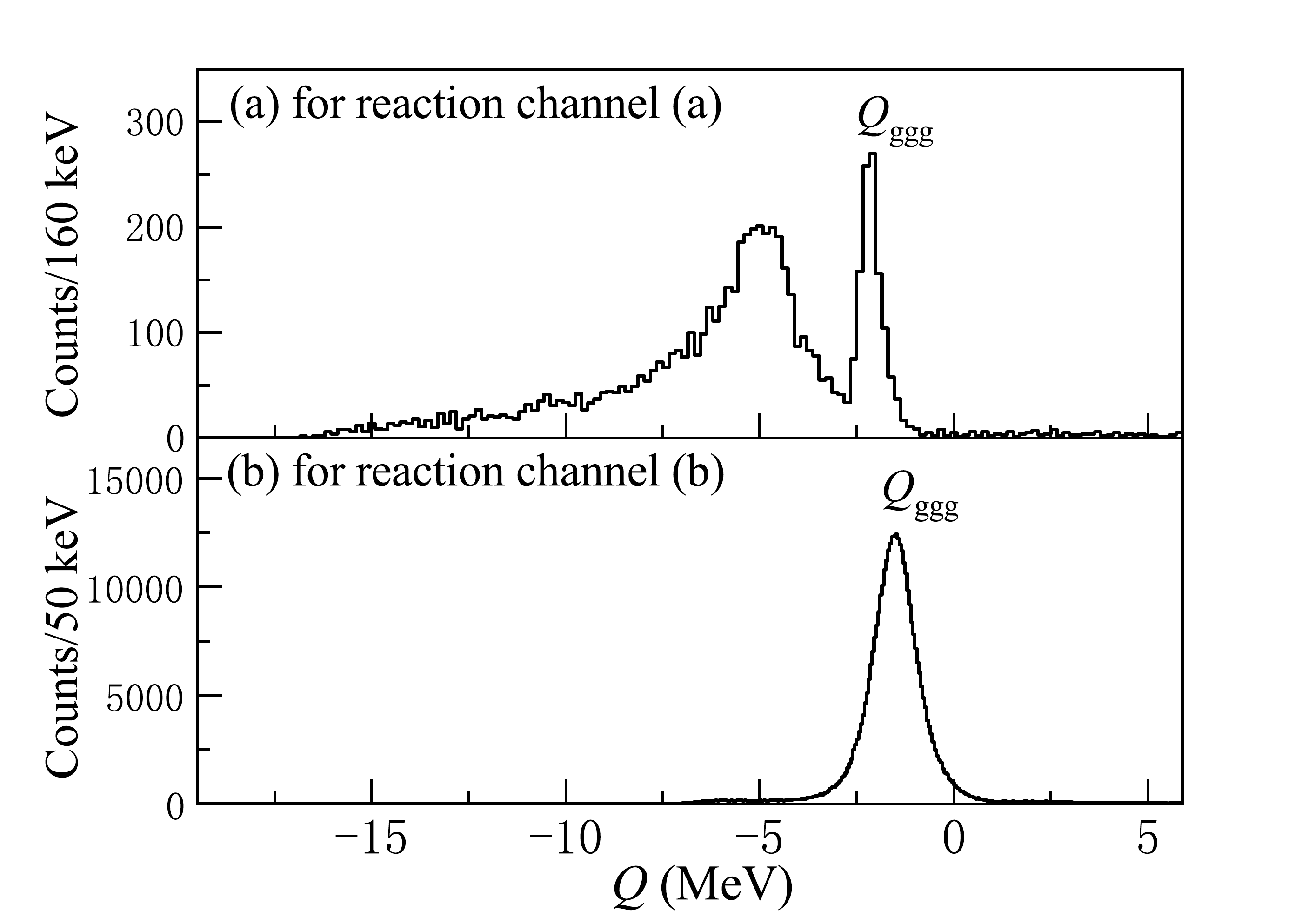}
\vspace{-4mm}
\caption{\footnotesize{Experimental $Q$-value spectra for the two reaction channels:
 (a) $\rm{^9Be}(\rm{^9Be},\rm{^{10}Be}^{*}\rightarrow\alpha + \rm{^6He})\rm{^{8}Be}$ ($Q_{\rm{ggg}}$ = -2.26 MeV); and (b) $\rm{^9Be}(\rm{^9Be},\rm{^{10}Be}^{*}\rightarrow\rm{n} + \rm{^9Be})\rm{^{8}Be_{g.s.}} \rightarrow 2\alpha$ ($Q_{\rm{ggg}}$ = -1.57 MeV).}}
\label{fig:2}
\end{center}
\vspace{-8mm}
\end{figure}

The energy released in a reaction, namely the reaction $Q$-value, is a useful quantity to select the reaction channels~\cite{Lijing2017}. It should be noted that the 7.54 MeV resonance in $^{10}$Be has only two possible particle-decay channels (n- and $\alpha$-decay), while its $\gamma$-decay is negligible ~\cite{Liendo2002}. The purpose of the present experiment is then to study two reaction-decay channels: (a)  \mbox{$\rm{^9Be}(\rm{^9Be},\rm{^{10}Be}^{*}\rightarrow\alpha + \rm{^6He})\rm{^{8}Be}$} \mbox{($Q_{\rm{ggg}}$ = -2.26 MeV)}; and (b) \mbox{$\rm{^9Be}(\rm{^9Be},\rm{^{10}Be}^{*}\rightarrow\rm{n} + \rm{^9Be})\rm{^{8}Be_{g.s.}} \rightarrow 2\alpha$ } \mbox{($Q_{\rm{ggg}}$ = -1.57 MeV)}. $Q_{\rm{ggg}}$ here means the $Q$-value for all particles in their ground states. The relative yields of these two channels, generated from the intermediate \mbox{7.54 MeV} resonance in $^{10}$Be, allow to deduce the absolute $\alpha$-decay BR of this state.

For channel (a), due to the interest on the low relative energy states, we only use events with $\alpha~+~\rm{^6{He}}$ pair being detected within one forward-angle telescope. The energy of the recoil ${^8}$Be can be deduced according tog momentum conservation. As a result,
$$ Q = E_{\rm{tot}}-E_{\rm{beam}}\nonumber = \Sigma{E_{i}}- E_{\rm{beam}},  \eqno{(1)} $$
where $E$ denotes the kinetic energy and $i$ runs for all particles in the exit channel. The corresponding spectrum is shown in Fig.~\ref{fig:2}(a), in which a narrow peak stands for $Q_{\rm{ggg}}$ and a broader one at lower $Q$ values is associated with the first excited state of $^8$Be (3.03 MeV, $2^+$).

\begin{figure}[!t]
\begin{center}
\includegraphics[width=.50\textwidth]{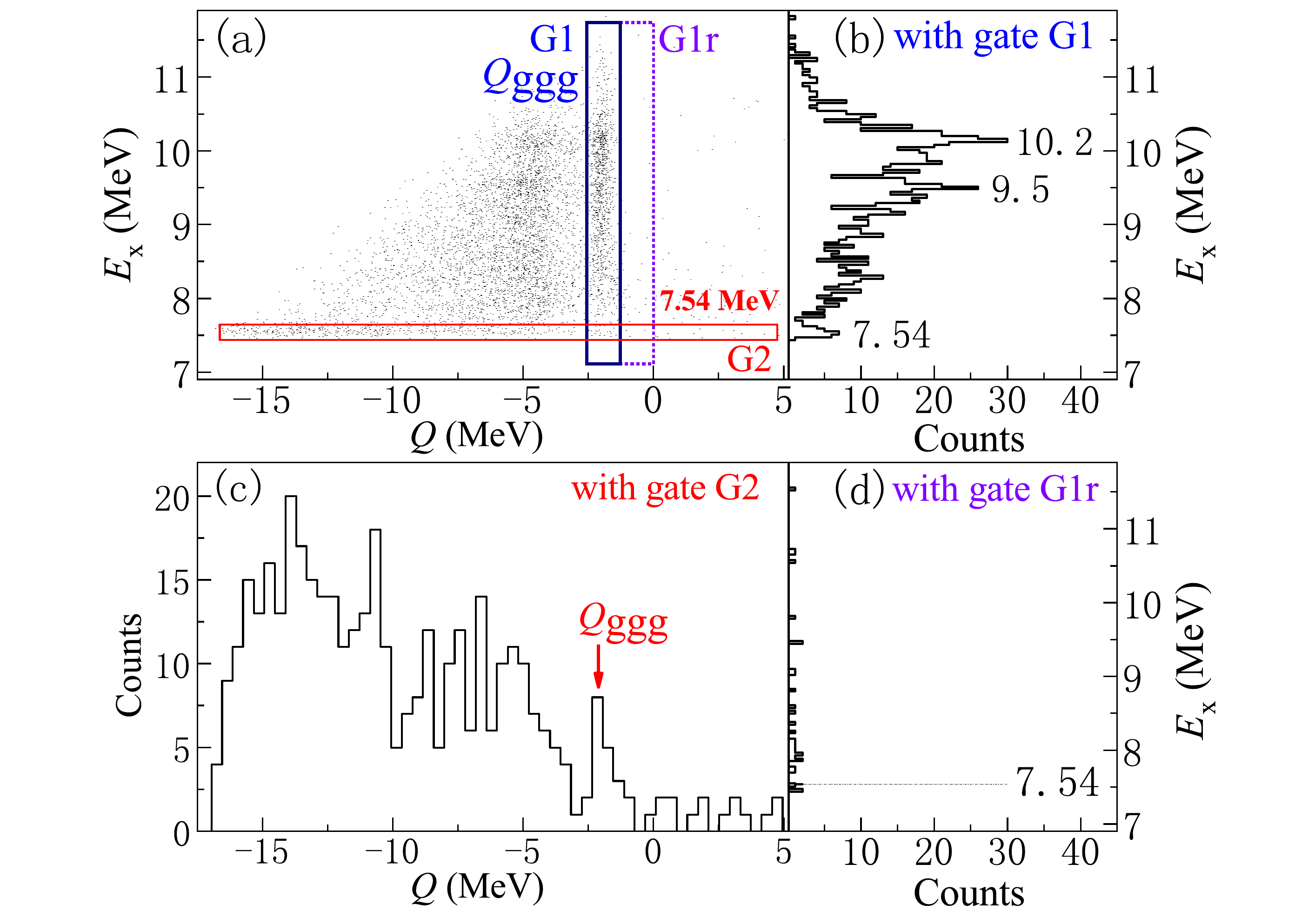}
\vspace{-6mm}
\caption{\footnotesize{Spectra deduced from the detected $\alpha$ + $^6$He pairs. (a) two-dimensional plot for $E_{\rm{x}}$ versus $\it{Q}$-value; (b) The projected $E_{\rm{x}}$ spectrum for $Q$-value around $Q_{\rm{ggg}}$ (gate G1); (c) The projected $Q$-value spectrum for $E_{\rm{x}}$ around 7.54 MeV (gate G2); (d) The projected $E_{\rm{x}}$ spectrum for $Q$-values at the right side of $Q_{\rm{ggg}}$ (gate G1r).}}
\label{fig:3}
\end{center}
\vspace{-10mm}
\end{figure}

The relative energy ($E_{\rm{rel}}$) of the decay products can be deduced according to the invariable mass (IM) method \cite{Yang2015}.  The associated excitation energy is thus $E_{\rm {x}} = E_{\rm{rel}} + E_{\rm{th}}$, with $E_{\rm{th}}$ the cluster separation energy \cite{Yang2015}. Since both $E_{\rm{x}}$ for $^{10}\rm{Be}$ and $\it{Q}$-value for the reaction channel (a) are calculable based on the detected $\alpha$ + $^6$He pair, the corresponding two-dimensional spectrum can be plotted to illustrate their possible correlations (Fig.~\ref{fig:3}(a)).

In order to have a strict constraint on the reaction mechanism, a gate on the $Q_{\rm{ggg}}$ peak (region G1 in Fig.~\ref{fig:3}(a)) is applied to the projection onto $E_{\rm{x}}$, as displayed in Fig.~\ref{fig:3}(b). As shown in the figure, the peak at around 9.5 and 10.1 MeV agree exactly with the previous observations \cite{Curtis2001,Curtis2004,Freer2006Be1015}, indicating the correctness of the present measurement and analysis.  But for the peak at about 7.54 MeV, very close to the threshold, cares must be  taken since there appears a relatively strong band at around this excitation energy but distributed broadly along the $Q$-dimension, as approximately indicated by the gate G2 in Fig.~\ref{fig:3}(a). To analyse its influence on the event counting for the 7.54 MeV resonance in $^{10}$Be, we project this band onto the $Q$-value dimension, as displayed in Fig.~\ref{fig:3}(c). It would be important to check the possible contamination to the $Q_{\rm{ggg}}$ peak by this background band. We find that this contamination depends quite sensitively on the PID selection. In our case, as demonstrated in Fig.~\ref{fig:1}, the identification for $^6$He is very clean, whereas that for $^4$He might be mixed by some nearby $^3$He.  We have plotted the $Q$-value spectrum gated on the left-side-half or right-side-half of the $^4$He peak (see the inset of Fig.~\ref{fig:1}), respectively. It was evidenced that the contamination to the $Q_{\rm{ggg}}$ peak is appreciable with the former gate, but negligible with the latter gate, similar to the background under the $Q_{\rm{ggg}}$ peak in Fig.~\ref{fig:2}(a) and Fig.~\ref{fig:3}(c), respectively. Therefore the latter selection of $^4$He, together with the normal selection of $^6$He, were adopted for the final analysis of the resonance, as presented in Fig.~\ref{fig:3}.  Of course, the efficiency simulation follows exactly the applied gates at the relevant steps. From Fig.~\ref{fig:3}(c) it can be seen that the $Q_{\rm{ggg}}$ peak is well distinguished from the background band distribution. The remaining minor scattered background exists all over the two-dimensional spectrum, as exhibited by Fig.~\ref{fig:3}(d), which can be naturally subtracted from the $E_{\rm{x}}$ spectrum.  Actually, from the well standing two-dimensional \mbox{$Q_{\rm{ggg}}$-$E_{\rm{x}}$(7.54 MeV)} peak (Fig.~\ref{fig:3}) and by subtracting the beneath background, we have obtained the counts for the resonance as $32 \pm 10$, taken into account the factor 2 for the $^4$He selection as mentioned above. The uncertainty here is statistical only, including the background contribution.

 We note that there are other two possible contaminating exit channels, namely $\rm{^{6}He + ^{12}C^* ( \rightarrow 3\alpha )}$ and $\alpha + \rm{^{14}C}^*$ $(\rightarrow \alpha + \rm{^{10}Be^*( \rightarrow \alpha + ^{6}He))}$, which are composed of the same mass combinations as the reaction channel (a). However, in the case of  $\rm{^{12}C^* \rightarrow 3\alpha}$ decay, the simulation shows that having one of these $\alpha$-particles going into the forward angle telescope while keeping another two closely (as $^8$Be g.s.) at large angle would require very high excitation energy in $^{12}$C ($> 10$ MeV). Furthermore, the Dalitz-plot for the possible $^{12}$C high-excitation versus the $^{10}$Be excitation, using the real data, does not show any structure correlation. Therefore, this background channel does not affect the actual extraction of the well-distinguished 7.54 MeV peak in $^{10}$Be. In the case of $^{14}$C production and decay, the two undetected $\alpha$-particles, one recoiling to a large angle and another from $^{14}$C-decay emitting to a forward angle, are separated from each other and cannot fake the $^8$Be(g.s.) as required by the actual $Q_{\rm{ggg}}$. The exclusion of this contamination has also been verified by simulation and real data analysis \cite{Lijing2017}. The effect of the target impurities, mainly carbon and oxygen contents, were analyzed and eliminated by using the EP-plot method ~\cite{Costanzo1990}. Event mixing was also checked against the $Q$-value spectrum.

To investigate the reaction channel (b), events were selected by requiring 2-particles being detected by a large-angle telescope (U1/D1) and one $^9$Be nucleus coincidentally detected by a forward-angle telescope (D0/U0). Again the energy of the missing neutron can be calculated according to the momentum conservation. Although the two particles detected by a large-angle telescope were not clearly identified due to their very low kinetic energies, the clear $Q_{\rm{ggg}}$ peak in Fig.~\ref{fig:2}(b) assures the case since any other mass combination must give much lower $Q$-value. This is further ascertained by gating on the relative energy of these two nearby particles as the g.s. of $^8$Be ($\sim$ 91 keV). Due to the larger uncertainties in detecting these two very low energy $\alpha$-particles the deduced $Q$-value spectrum exhibits a larger peak-width (Fig.~\ref{fig:2}(b)). Remarkably, there is almost no continuous background in the lower $Q$-value region, owing to the three-folds coincident detection.

\begin{figure}[!b]
\vspace{-4mm}
\begin{center}
\includegraphics[width=.49\textwidth]{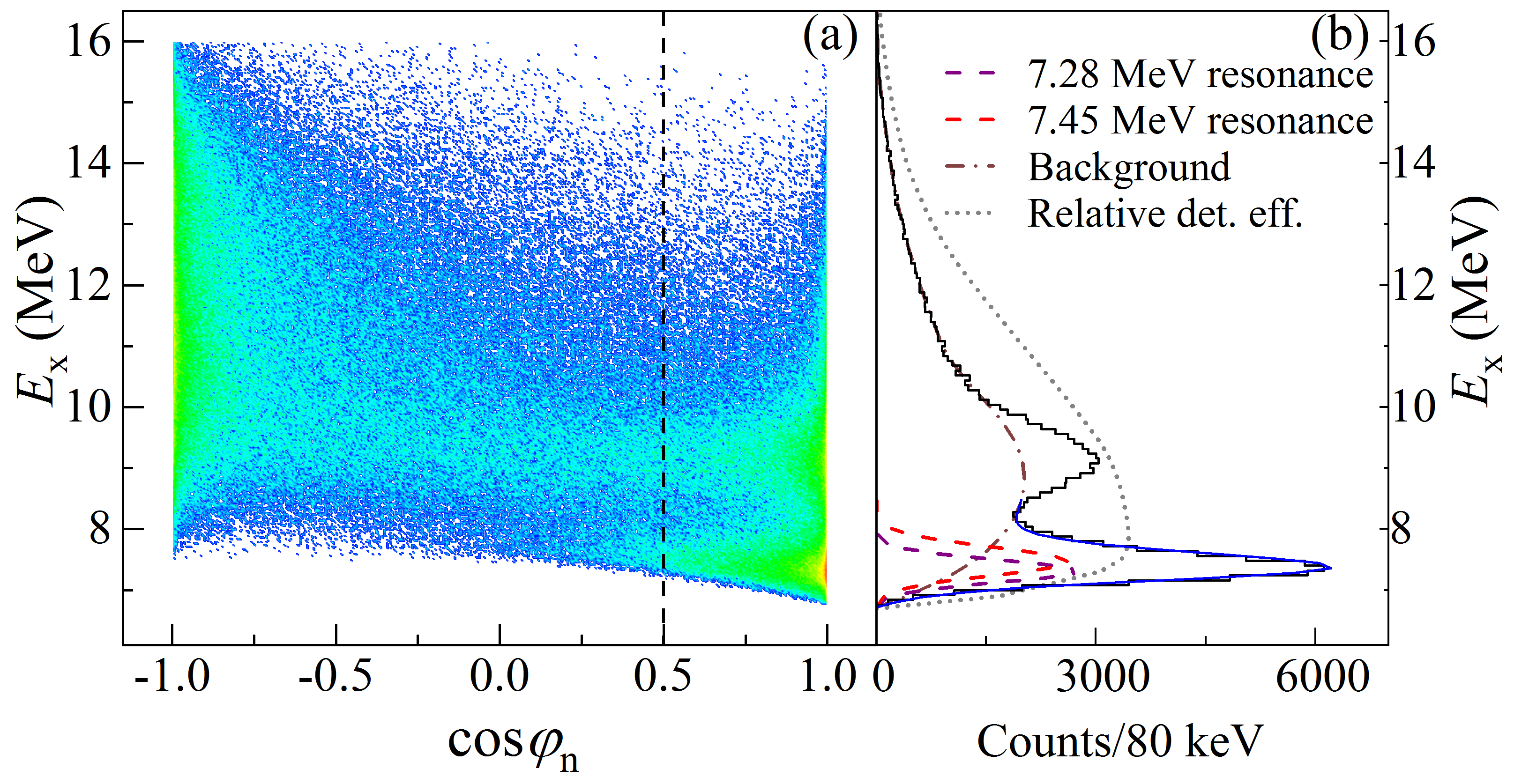}
\vspace{-7mm}
\caption{\footnotesize{(a) Two-dimensional plot for the reconstructed $E_{\rm{x}}$($\rm{^{10}Be}$) versus cos$\varphi_{\rm{n}}$. $\varphi_{\rm{n}}$ stands for the azimuthal angle of the deduced neutron, as defines in the text. (b) Excitation energy $E_{\rm{x}}$ in $^{10}$Be deduced from the $\rm{n}$ + $^9$Be decay channel, subject to event selection as described in the text. The curves are also explained in the text.}}
\label{fig:4}
\end{center}
\vspace{-5mm}
\end{figure}

The presently targeted exit channel is inevitably accompanied by a very probable inelastic scattering channel $\rm{^9Be}(\rm{^9Be},\rm{^{9}Be}^{*}\rightarrow\rm{n}+\rm{^8Be})\rm{^{9}Be}$, which possesses the same $Q$-value but does not reflect the $^{10}$Be excitation. In order to reduce this contamination, $E_{\rm{x}}$ from $^9$Be + n reconstruction is plotted against cos$\varphi_{\rm{n}}$ , as shown in Fig.~\ref{fig:4}(a). Here $\varphi_{\rm{n}}$ stands for the azimuthal angle of the deduced neutron in the laboratory system, with the 0-degree axis lies in the horizontal plan and points to the side having the $^9$Be detection. As presented in Fig.~\ref{fig:4}(a), cos$\varphi_{\rm{n}}$ distribution is concentrated around +1 and -1, with the real resonances in $E_{\rm{x}}$, such as the one at about 7.5 MeV, placed at the +1 end. This plot provides a good discrimination between the targeted reaction channel (b) and the above mentioned contamination channel. In fact, the former tends to emit a neutron close by the forward moving $^9$Be, while the latter combined with the recoil $^8$Be at the opposite side of $^9$Be. Considering both the signal to background ratio and the signal detection efficiency, we require $\cos\varphi_{\rm{n}} \geq 0.5$ when accumulating the $E_{\rm{x}}$ spectrum for $^{10}$Be (Fig.~\ref{fig:4}(b)). We note that Fig.~\ref{fig:4}(a) could be replaced by the \mbox{Dalitz-plot} of $^{10}\rm{Be}^*$ versus $^{9}\rm{Be}^*$. However the present method provides a better performance.

Previously the excited states in $^{10}$Be were observed at 7.37 and 7.54 MeV \cite{complication2004}, which should be included in our fitting analysis of the relatively broad peak around 7.5 MeV. The detection efficiency and energy resolution as a function of $E_{\rm{x}}$ were estimated by Monte Carlo simulation taking into account the realistic detector setup and performances ~\cite{Yang2014-b,Hai2008}. Dotted line in Fig.~\ref{fig:4}(b) show the the relative efficiency. The resolution ranges from $\sim$200 keV (standard deviation $\sigma$) at $E_{\rm{x}}$ = 7.37 MeV up to $\sim$240 keV at $E_{\rm{x}}$ = 7.54 MeV. The intrinsic widths of the resonances were previously reported as $\sim$10 keV for both states ~\cite{complication2004}. The data after efficiency correction were fitted by using two Gaussian-shape functions  with a fixed interval of 170 keV between their central positions (dashed lines in Fig.\ref{fig:4}(b)), together with a smooth background function ~\cite{Hai2008} (dot-dashed line in Fig.\ref{fig:4}(b)). The adopted widths ($\sigma$) after the variation are \mbox{195 keV} and 245 keV for the 7.37 MeV and 7.54 MeV peaks, respectively, being consistent with the simulation results. Finally, the counts $N_{\rm{n}} = 19146\pm 138$ is determined for the decay $\rm{^{10}Be}(7.54~MeV) \rightarrow\rm{n}+\rm{^9Be}$ , subject to the applied cuts which will be accounted for by the efficiency simulation. The  uncertainty here is statistical only. By using events in this dominating neutron decay channel, together with the integrated incident particle number, the target thickness and the simulated detection solid angle, we obtain a population cross section of 65 $\mu$b/sr at about 25$^0$ lab. (15 $\mu$b/sr at about 55$^0$ c.m.) for the 7.54 MeV state in $^{10}$Be. Here the statistical uncertainty is less than 1$\%$ while the systematic uncertainty is estimated to be about 10$\%$ resulted mainly from the beam integration.

Monte Carlo simulations were conducted for both reaction-decay channels (a) and (b), using realistic experimental setup and event selection configurations. The differential cross section for transferring into the \mbox{7.54 MeV ($2^+_3$)} state in $^{10}$Be was generated by the Distorted-Wave-Born-Approximation (DWBA) calculation using the code Fresco \cite{Fresco1988}. The optical potential parameters were taken from Refs.~\cite{9be+9bepotential,8Be+npotential}. From the simulation the ratio of the acceptances for the two reaction-decay channels,  $\varepsilon_{\alpha}$/$\varepsilon_{\rm{n}}$,  is determined to be 4.14.

The BR for $\alpha$-decay from $^{10}$Be(7.54 MeV) is now expressed in the form:
$$\frac{\Gamma_{\alpha}}{\Gamma_{\rm{tot}}}= \frac{N_{\alpha}/\varepsilon_{\alpha}}{N_{\alpha}/\varepsilon_{\alpha}+N_{\rm{n}}/\varepsilon_{\rm{n}}}= \frac{N_{\alpha}}{N_{\alpha}+N_{\rm{n}}\frac{\varepsilon_{\alpha}}{\varepsilon_{\rm{n}}}} \eqno{(2)}$$
where $N_{\alpha}$ and $N_{\rm{n}}$ represent the detected $\alpha$ and neutron numbers from the resonance. Using the known numbers we obtain $\Gamma_{\alpha}$/$\Gamma_{\rm{tot}}$ = $(4.04\pm1.26)\times 10^{-4}$. The error here is statistical only. In addition some systematic error of about $10\%$ is estimated, considering the reasonable parameter variations in the simulation and the function selection in the fitting procedure. The presently obtained cluster-decay BR is significantly smaller than the previously measured  ones \cite{Liendo2002,Milin2005}. The existing observed BR$_\alpha$ results are listed in Table~\ref{BR}.
\begin{table}[!b]
\vspace{-7mm}
\caption{Summary of the $\alpha$-decay branching ratio (BR$_\alpha$) of the 7.54 MeV ($2_3^+$) state in $^{10}$Be. \label{BR}}
\begin{tabular*}{\hsize}{@{}@{\extracolsep{\fill}}ccc@{}}
\hline
   Data source  &  BR$_\alpha$  \\
\hline
Experimental result from ~\cite{Liendo2002}                        & $(3.5 \pm 1.2)\times 10^{-3}$    \\
Experimental result from ~\cite{Milin2005}                         & $\geq (2.0 \pm 0.6)\times 10^{-3}$  \\
Present experimental result                                        & $(4.04 \pm 1.26)\times 10^{-4}$     \\
Converted from $^{10}$B analog$^a$           &  $(1.3 \pm 0.3)\times 10^{-4}$           \\
\hline
\end{tabular*}
\footnotetext[1]{adapted from Ref.~\cite{Kuchera2011} for a reduced radius $r_0$ = 1.4 fm.}
\end{table}

Based on the single-channel $R$-matrix approach, the $\alpha$-decay partial width $\Gamma_{\alpha}$ can be converted into the reduced width $\gamma^2_{\alpha}$ according to the formula~\cite{Yang2014-b,Rmatrix1958}:
$$\gamma^2_{\alpha}= \frac{\Gamma_{\alpha}}{2P_l} , \eqno{(3)}$$
with $P_l$ being the barrier penetrability factor \cite{Yang2014-b,Rmatrix1958}. $\gamma^2_{\alpha}$ is generally presented with respective to the Wigner Limit $\gamma^2_{\alpha W}$, leading to the dimensionless reduced width $\theta^2_{\alpha}$ = $\gamma^2_{\alpha}$/$\gamma^2_{\alpha W}$ with
$\gamma^2_{\alpha W}$ = $3\hbar^2$/($2\mu R^2$) ~\cite{Freer2006Be1015, Descouvemont2010}. Here $\mu$ is the reduced mass of the decaying cluster system and $R$ the channel radius given by $R=r_0(A_1^{1/3}+A_2^{1/3})$.
$\theta^2_{\alpha}$ is also interpreted as the ``$\alpha$-cluster spectroscopic factor (SF)''~\cite{Freer2006Be1015}, which is sensitively dependent on $R$. As a matter of fact, extremely large deformation or inter-$\alpha$-core distance has been predicted for the $\sigma$-bond molecular states in $^{10}$Be, in comparison to its ground state \cite{Horiuchi2012,Enyo1999}. Using the known total width $\Gamma_{\rm{tot}} = 6.3 ~\rm{keV}$ \cite{Fortune2011Be10} and BR$_\alpha$ measured in the present experiment, we obtain $\theta^2_{\alpha}$ ranging from $2.56(80)$ to 0.87(27) for the reduced radius $r_0$ from 1.4 to 1.8 fm. This SF$_\alpha$ result is similar to that for the $4_1^+$ state in the same molecular rotational band, which ranges from 2.23 to 0.66 for $r_0$ from 1.4 to 1.8 fm ~\cite{Freer2006Be1015}, and is fairly consistent with the maximum degeneracy of $\alpha$-particle.

We notice that the cluster SF for $2_3^+$ isobaric analog state in $^{10}$B (8.894 MeV) was reported to be 0.73(13)~\cite{Kuchera2011}, when deduced by using the standard reduced channel radius $r_0$ = 1.4 fm and the Wigner Limit as defined above ~\cite{Freer2006Be1015, Descouvemont2010}. The difference between this value and our results for $^{10}$Be (7.54 MeV) is twice as large as the summed error bar. The same difference is also evidenced by the BR$_\alpha$ in Table~\ref{BR}. This kind of cluster-SF difference between analog states was observed at other occasions as well. For instance, the 8.898 MeV (3$^-$; $T$ = 1) state in $^{10}$B has a reported large SF$_\alpha$ of 0.42, whereas its analog in $^{10}$Be (7.31 MeV, 3$^-$ state) is a well-recognized pure single-particle state ~\cite{Kuchera2011}. Indeed, theoretical studies using the Gamow shell model have revealed that, for the weakly bound or unbound systems, the structure of isobaric analog states varies within the isomultiplet and impacts the associated particle SF~\cite{Michel2010}. This variation is mainly related to the large asymmetry between proton and neutron emission thresholds which modify the respective coupling to the continuum. Since the cluster formation occurs most likely at around the cluster separation threshold and is often mixed with single-particle configuration especially in the case of unstable nuclei, the spectroscopic change within the clustering isomultiplet might be enhanced. As an example, this kind of structure change has been demonstrated by the $^{10}$Be - $^{10}$C mirror system. The allowed $2p$-emission from the $0_2^+$ or $2_3^+$ level of $^{10}$C leads to significant structure change in comparison to its mirror counterpart in $^{10}$Be which has no corresponding $2n$-emission channel because of the relatively higher threshold~\cite{Golderg2012}. In addition, the isospin mixing may also result in spectroscopic change. For example, the strong clustering $2^+$(8.894 MeV; $T = 1$) state in $^{10}$B has a reported isospin-conservation $\alpha$-decay width of about 18 keV, together with an isospin-violating decay width of about 12 keV~\cite{Kuchera2011}. This isospin-mixing in decay process does not exist in the $^{10}$Be analog state (7.54 MeV). We notice again that the extraction of the cluster SF depends sensitively on the channel radius of the resonance, which may be changed state by state for weakly bound or unbound system, as recently demonstrated in Ref.~\cite{Ito2018}. Thus, the valuable comparison among SFs of the analog states requires also independent determination of the radius. It is obvious that the precise and independent measurements of cluster-decay BRs and other observables for analog states would provide important information for the study of isospin-symmetry breaking in exotic composite nuclear systems.

In summary, a new experimental investigation of the cluster structure of the 7.54 MeV (2$^+$) resonant state in $^{10}$Be was performed by using the reaction \mbox{$\rm{^9Be}$ $(\rm{^9Be},\rm{^{10}Be})\rm{^{8}Be}$} at 45 MeV beam energy. Special measures were taken to reduce the strong near-threshold background and to assure a reliable extraction of the cluster-decay events from the resonance. The neutron-decay from the same resonance was also analyzed based on the three-fold coincident measurement. A cluster-decay branching ratio of $(4.04\pm1.26)\times 10^{-4}$ is determined for the 7.54 MeV resonance in $^{10}$Be. The deduced $\alpha$-cluster SF is from 2.56(80) to 0.87(27) for the reduced channel radius $r_0$ from 1.4 to 1.8 fm. The present work, together with the well established cluster-description of the 6.18 MeV ($0^+$) and 10.15 MeV ($4^+$) states, leads to a comprehensive understanding of a perfect molecular rotational band in excited $^{10}$Be. The comparison between the currently obtained cluster SF with that of the analog state in $^{10}$B may stimulate further studies of the isospin-symmetry breaking in clustering nuclear systems.\\

\begin{acknowledgments}
The authors wish to thank the staffs of the tandem accelerator laboratory at CIAE for their excellent work in providing the beams. This work is support by the National Key R$\&$D Program of China
(Grant No. 2018YFA0404403) and the National Natural Science Foundation of China (Grant Nos. 11535004, 11875074, 11875073, U1967201, 11635015, 11775004, 11775013 and 11775316).
\end{acknowledgments}

\begin{thebibliography}{99}


\bibitem{Ikeda1968} K. Ikeda, N. Tagikawa, and H. Horiuchi, Prog. Theor. Phys. Suppl. 68 (1968) 464.
\bibitem{Oer2006} W. von Oertzen, M. Freer, and Y. Kanada-En'yo, Phys. Rep. 432 (2006) 43.
\bibitem{Horiuchi2012} H. Horiuchi, K. Ikeda, and K. Kat, Prog. Theor. Phys. Suppl. 192 (2012) 1.
\bibitem{Lijing2017}
J. Li, Y. L. Ye, Z. H. Li $et$ $al$., Phys. Rev. C 95 (2017) 021303(R).
\bibitem{Freer2018}
M. Freer and H. Horiuchi and Y. Kanada-En¡¯yo and D. Lee, Rev. Mod. Phys. 7 (2018) 035004.
\bibitem{Yang2014}
Z. H. Yang, Y. L. Ye, Z. H. Li $et$ $al$., Phys. Rev. Lett. 112 (2014) 162501.
\bibitem{Buck1977}
B. Buck, H. Friedrich, C. Wheatley, Nucl. Phys. A 275 (1977) 246.
\bibitem{Oer1996}
W. von Oertzen, Z. Phys. A 354 (1996) 37.
\bibitem{Oer1997}
{W. von Oertzen, Z. Phys. A 357 (1997) 355}.
\bibitem{Enyo1999}
{Y. Kanada-En'yo, H. Horiuchi, and A. Dot\'e, Phys. Rev. C 60 (1999) 064304}.
\bibitem{Lyu2016Be10}
{M. Lyu, Z. Ren, B. Zhou $et$ $al$., Phys. Rev. C 93 (2016) 054308}.
\bibitem{Fortune2011Be10}
{H. T. Fortune and R. Sherr, Phys. Rev. C 84 (2011) 024304}.
\bibitem{Aquilla2016}
{D. Dell' Aquila, I. Lombardo, L. Acosta $et$ $al$., Phys. Rev. C 93 (2016) 024611}.
\bibitem{Kra2017}
{K. Kravvaris and A. Volya, Phys. Rev. Lett. 119 (2017) 062501}.
\bibitem{Lyu2018}
{M. Lyu and K. Yoshida and Y. Kanada-En'yo and K. Ogata, Phys. Rev. C 97 (2018) 044612}.
\bibitem{Freer2006Be1015}
{M. Freer $et$ $al$., Phys. Rev. Lett. 96 (2006) 042501}.
\bibitem{Suzuki2013}
{D. Suzuki $et$ $al$., Phys. Rev. C 87 (2013) 054301}.
\bibitem{Liendo2002}
{J. A. Liendo, N. Curtis, D. D. Caussyn, N. R. Fletcher, and T. Kurtukian-Nieto, Phys. Rev. C 65 (2002) 034317}.
\bibitem{Milin2005}
{M. Milin $et$ $al$., Nucl. Phys. A 753 (2005) 263}.
\bibitem{ANDERSON1974}
{R. E. Anderson $et$ $al$., Nucl. Phys. A 236 (1974) 77}.
\bibitem{Kuchera2011}
{A. N. Kuchera $et$ $al$., Phys. Rev. C 84 (2011) 054615}.
\bibitem{Carbone2014}
{D. Carbone $et$ $al$., Phys. Rev. C 90 (2014) 064621}.
\bibitem{Harakeh1980}
{M.N. Harakeh $et$ $al$., Nucl. Phys. A 344 (1980) 15}.
\bibitem{tianzy2016}
{Z. Y. Tian, Y. L. Ye, Z. H. Li $et$ $al$., Chin. Phys. C 40 (2016) 111001}.
\bibitem{Jiang2017}
{W. Jiang, Y. L. Ye, Z. H. Li $et$ $al$., Sci. China: Phys. Mech. Astron. 60 (2017) 062011}.
\bibitem{Qiaorui2014}
{R. Qiao, Y. L. Ye, J. Wang, Z. H. Li, H. B. You, Z. H. Yang, and B. Yang, IEEE Tran. Nucl. Sci. 61 (2014) 596}.
\bibitem{CalibrationMilin2014}
{M. Uroi\'c $et$ $al$., Eur. Phys. J. A 51 (2015) 93}.
\bibitem{Freer2004BeB}
{S. Ahmed $et$ $al$., Phys. Rev. C 69 (2004) 024303}.
\bibitem{Yang2015}
{Z. H. Yang, Y.L. Ye, Z.H. Li $et$ $al$., Phys. Rev. C 91 (2015) 024304}.
\bibitem{Curtis2001}
{N. Curtis $et$ $al$., Phys. Rev. C 64 (2001) 044604}.
\bibitem{Curtis2004}
{N. Curtis $et$ $al$., Phys. Rev. C 70 (2004) 014305}.

\bibitem{Costanzo1990}
{E. Costanzo $et$ $al$., Nucl. Instr. Meth. A 295 (1990) 373}.
\bibitem{complication2004}
{D. R. Tilley $et$ $al$., Nucl. Phys. A 745 (2004) 155}.
\bibitem{Yang2014-b}
{Z. H. Yang, Y. L. Ye, Z. H. Li $et$ $al$., Sci. China: Phys. Mech. Astron. 57 (2014) 1613}.
\bibitem{Hai2008}
{P. J. Haigh, N.I. Ashwood, T. Bloxham $et$ $al$., Phys. Rev. C 78 (2008) 014319}.
\bibitem{Fresco1988}
{I. J. Thompson, Comput. Phys. Rep. 7 (1988) 167}.
\bibitem{9be+9bepotential}
{A. R. Omar, J. S. Eck, J. R. Leigh, and T. R. Ophel, Phys. Rev. C 30 (1984) 896}.
\bibitem{8Be+npotential}
{L. Jarczyk $et$ $al$., J. of Phys. G: Nucl. Phys. 11 (1985) 843}.
\bibitem{Rmatrix1958}
{A. M. Lane and R. G. Thomas, Rev. Mod. Phys. 30 (1958) 257}.
\bibitem{Descouvemont2010}
{P. Descouvemont and D. Baye, Rep. Prog. Phys. 73 (2010) 036301}.
\bibitem{Michel2010}
{N. Michel, W. Nazarewicz and M. Ploszajczak, Phys. Rev. C 82 (2010) 044315}.
\bibitem{Golderg2012}
{V. Z. Goldberg and G. V. Rogachev, Phys. Rev. C 86 (2012) 044314}.
\bibitem{Ito2018}
{M. Ito, Phys. Rev. C 97 (2018) 044608}.

\end{thebibliography}

\end{document}